# Provincial allocation of China's commercial building operational carbon towards carbon neutrality

Yanqiao Deng [1], Minda Ma [2 *, 3], Nan Zhou [3], Chenchen Zou [1], Zhili Ma [1], Ran Yan [1], Xin Ma [4]

1. School of Management Science and Real Estate, Chongqing University, Chongqing, 400045, PR China
2. School of Architecture and Urban Planning, Chongqing University, Chongqing, 400045, PR China
3. Building Technology and Urban Systems Division, Energy Technologies Area, Lawrence Berkeley National Laboratory, Berkeley, CA 94720, United States
4. School of Mathematics and Physics, Southwest University of Science and Technology, Mianyang, 621010, PR China

- Corresponding author: Dr. Minda Ma, Email: maminda@lbl.gov

  Homepage: https://buildings.lbl.gov/people/minda-ma

  https://chongjian.cqu.edu.cn/info/1556/6706.htm

## Graphical Abstract

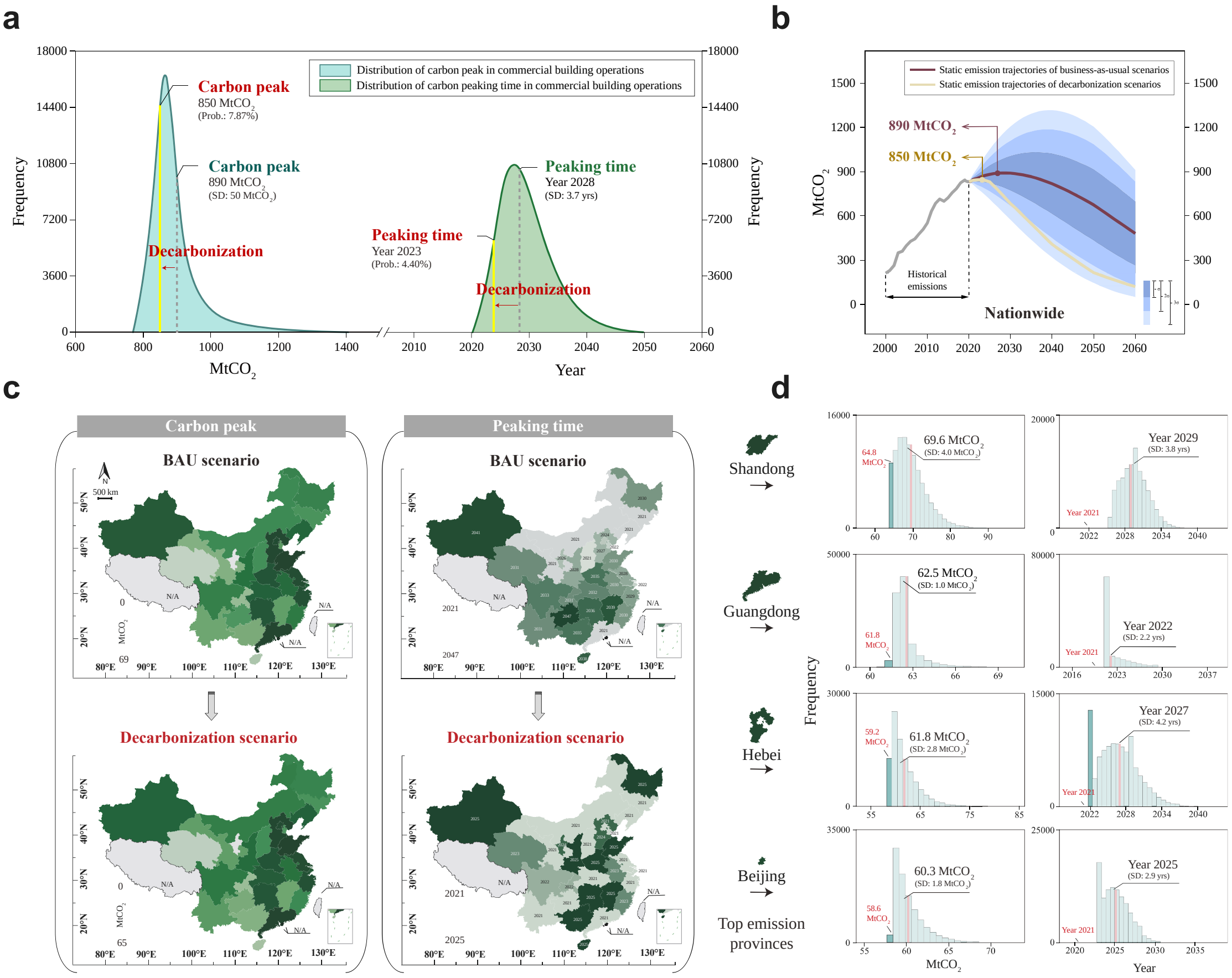

**Graphical abstract:** (a) Distributions of the carbon peak and peaking time for China's commercial building operations nationwide; (b) National emission trends and projections from 2000 to 2060; (c) Provincial distributions of carbon peaks and the corresponding peaking time under BAU and decarbonization scenarios; (d) Carbon peaks and peaking time for the four provinces with the highest emission levels.

## Highlights

- National carbon peak and provincial carbon allocation are optimized by the dynamic scenario analysis.
- Operational commercial building emissions in China will peak in 2028 (± 3.7 yrs) at 890 (± 50) $MtCO_2$.
- Emissions across provinces are expected to peak before 2046 (± 4.8 yrs) under dynamic scenario analysis.
- The future emission peak in Shandong [69.6 (± 4.0) $MtCO_2$)] will be 11 times greater than that in Ningxia.
- East China's reduction allocation of 18.1 $MtCO_2$ will exceed the lowest emission regions by 6.7 times.

## Abstract

National carbon peak track and optimized provincial carbon allocations are crucial for mitigating regional inequality within the commercial building sector during China's transition to carbon neutrality. This study proposes a top-down model to evaluate carbon trajectories in operational commercial buildings up to 2060. Through Monte Carlo simulation, scenario analysis is conducted to assess carbon peak values and the corresponding peaking year, thereby optimizing carbon allocation schemes both nationwide and provincially. The results reveal that (1) the nationwide carbon peak for commercial building operations is projected to reach 890 (± 50) megatons of carbon dioxide ($MtCO_2$) by 2028 (± 3.7 years) in the case of the business-as-usual scenario, with a 7.87% probability of achieving the carbon peak under the decarbonization scenario. (2) Significant disparities will exist among provinces, with Shandong's carbon peak projected at 69.6 (± 4.0) $MtCO_2$ by 2029, approximately 11 times higher than Ningxia's peak of 6.0 (± 0.3) $MtCO_2$ by 2027. (3) Guided by the principle of maximizing the emission reduction potential, the optimal provincial allocation scheme reveals the top three provinces requiring the most significant reductions in the commercial sector: Xinjiang (5.6 $MtCO_2$), Shandong (4.8 $MtCO_2$), and Henan (4.7 $MtCO_2$). Overall, this study offers optimized provincial carbon allocation strategies within the commercial building sector in China via dynamic scenario simulations, with the goal of hitting the carbon peak target and progressing toward a low-carbon future for the building sector.

## Abbreviation notation

BAU – Business-as-usual

Mtce – Megatons of coal equivalent

$MtCO_2$ – Megatons of carbon dioxide

SD – Standard deviations

## Nomenclature

$C_c$ – Total operational carbon emissions of commercial buildings

$e_c$ – Energy intensity of commercial buildings

$f_c$ – Per capita floor area of commercial buildings

$K_c$ – Total carbon emission factor of commercial buildings

$K_{coal}$ – Carbon emission factor of coal

$K_{electricity}$ – Carbon emission factor of electricity generation

$K_{gas}$ – Carbon emission factor of natural gas

$\omega_i$ – Random parameter of model parameter $i$

$P$ – Population size

$R_{c-BIPG}$ – Proportion of self-generated energy in commercial buildings

$R_{c-coal}$ – Proportion of coal used in commercial building energy use

$R_{c-electricity}$ – Electrification rate of commercial buildings

$R_{c-gas}$ – Proportion of natural gas used in commercial building energy use

## 1. Introduction

Buildings accounted for 34% of global energy demand and 37% of corresponding carbon dioxide ($CO_2$) emissions in 2022, significantly contributing to global climate change [1]. The combined operations of commercial and residential buildings generated more than one-third of global energy use and greenhouse gas emissions [2]. In China, transitioning the building sector to a low-carbon scenario is crucial for bridging the carbon mitigation gap and enabling the country to reach its carbon peak goal ahead of 2030 [3]. To achieve this goal, it is essential to allocate national carbon emission control targets to each province. Traditional "one-size-fits-all" and locally self-declared task allocation methods are insufficient for the provincial distribution of building carbon emissions and often fail to align with scientific decision-making principles [4]. Therefore, optimized carbon allocation schemes for China's provincial commercial buildings are vital for effectively progressing towards carbon neutrality in buildings by the year 2060.

An increasing number of studies have extensively examined historical energy use and carbon emissions [5, 6], prospective future emission trends [7], and potential emission reduction assessments [8] in China's commercial building operations, offering guidance for controlling carbon emissions and reducing regional disparities. Nevertheless, significant regional heterogeneity across provinces [9], such as variations in population scales, urbanization stages, economic growth levels, and energy structure characteristics [10], leads to uneven spatial-temporal distributions of carbon emissions [11, 12]. The influence of these variations on national emission changes and carbon peak targets remains unclear, contributing to challenges in implementing national emission reduction targets at the provincial level [13]. Some studies have explored regional decarbonization pathways and emission allocation schemes [14, 15], such as scenario analyses of green and low-carbon development in Shaanxi province [16]. However, conventional fixed allocation methods commonly used in these studies may fail to adequately address the complexities of China's commercial building operations [17]. Moreover, research on optimizing dynamic carbon peaks and allocation schemes across provinces with scenario analysis of commercial buildings is relatively rare. The following issues are covered in this study to help fill these gaps.

- How do operational carbon trajectories evolve in commercial buildings nationwide?
- How do carbon emission peaks vary across provinces in commercial building operations?

- How to allocate carbon budget for each province based on decarbonization potential maximum?

To address these challenges, an operational emission model for commercial buildings is built from the top-down perspective, integrating business-as-usual (BAU) and decarbonization scenarios to assess carbon emissions and energy use in the Chinese commercial building operations up to the year 2060. Subsequently, dynamic scenario analyses are conducted via Monte Carlo simulations to calculate the future carbon emission peak probabilities at the provincial level. Finally, guided by the principle of maximizing the emission reduction potential, an optimized allocation scheme for provincial carbon emissions is proposed, aiming to promote operational decarbonization within commercial buildings.

**The primary contribution of our study** is the development of a dynamic carbon allocation method, specifically designed for provincial commercial building operations in China, through Monte Carlo simulation that accounts for uncertainties, thereby optimizing provincial carbon allocation based on the decarbonization potential maximum. By evaluating the dynamic carbon trajectory of commercial buildings, this work conducts dynamic scenario analyses on operational carbon emission peaks and peaking time across provinces and further optimizes the provincial carbon emission allocation. This work offers valuable insights into the provincial-level distribution of carbon emissions and accelerates efforts to realize net-zero vision in buildings by the mid-century.

The remaining structure of this study is organized as follows: Section 2 presents a recent literature review. Section 3 provides the methodological framework and data collection. Section 4 shows the static carbon emission trajectories and scenario analysis under the BAU and decarbonization scenarios. Section 5 further explores dynamic carbon emission peak simulations, incorporates uncertainty considerations, and discusses provincial carbon peaks and optimized allocation strategies. Finally, Section 6 presents the key conclusions of this study.

## 2. Literature review

Forecasting operational carbon emission peaks in buildings generally involves the bottom-up and top-down approaches. The bottom-up approach estimates total provincial or national building carbon emissions and energy use by aggregating data, which often integrates platform models for low emissions analysis [18, 19], decomposition methods [20], system dynamics models [21, 22], etc., combined with scenario analysis for accurate carbon emission prediction. Conversely, the top-down approach in the building energy sector analyzes socioeconomic factors, including urbanization rate [23], population [24], and per capita floor area [25], to assess overall trends. While this approach offers fewer technical details, it focuses on observed macroeconomic trends [26]. Common models include the equation of Human Impact, Population, Affluence, and Technology model series [27] (such as the Kaya identity [28, 29] and the Stochastic Impacts by Regression on Population, Affluence, and Technology model [30], the environmental Kuznets curve [31], etc. These models, when applied to the commercial building, primarily assess carbon peaks nationwide or focus on a limited number of provinces. However, under the national carbon neutrality target for 2060, significant spatial heterogeneity in emission reduction allocations persists across regions [32]. Therefore, relying solely on static bottom-up or top-down approaches to evaluate carbon emission peaks and provincial allocations is insufficient [33, 34].

The scenario analysis method has been extensively used in the building sector to simulate developmental changes and forecast potential future pathways for carbon emission reduction across various scenarios [35]. Several studies have employed bottom-up or top-down models with scenario analysis to analyze carbon peaks [16, 36]. For example, Zhang et al. [37] adjusted essential indicators and established distinct scenarios to simulate the carbon peaks in China. However, most existing studies lack consideration of the uncertainty of parameters in commercial building operations, potentially leading to impractical energy savings and emission reduction goals. A growing number of studies have adopted dynamic scenario analysis for energy and operational carbon emission reduction, integrating methods such as the Monte Carlo simulation [38, 39], system dynamics method [40, 41], dynamic material flow principle [5], etc. Li et al. [12] used the Monte Carlo simulation to investigate potential dynamic carbon emission peaks, incorporating uncertainties in key parameters. Their findings indicated that energy use in commercial buildings significantly contributed to emission growth and most provinces were projected to fall short of meeting their emission peak targets.

These models, along with scenario analysis, have been used in previous studies on building carbon emissions to assess carbon emission peaks, providing valuable insights for the current research. However, there are two gaps that still require further investigation.

With respect to the research topic, most studies on carbon allocation in commercial building operations have focused primarily on the overall regional allocation of carbon emissions [42, 43]. However, few studies have evaluated the carbon allocation for each province while considering national carbon emission peak controls. In addition, Joensuu et al. [44] pointed out that mainstream allocation techniques, such as zero-sum game theory [45], the bankruptcy game model [46], and data envelopment analysis [15], are unsuitable for application in the building sector. To date, only a limited number of studies have explored optimal carbon allocation strategies for China's commercial building operations at the provincial level.

With respect to the methodology, few studies have integrated the scenario analysis using Monte Carlo approach and the Kaya identity within a top-down framework to assess the operational carbon peaks of commercial buildings both nationwide and provincially. Instead of relying on static scenario analysis, which overlooks the uncertainties of key parameters, Monte Carlo simulation enables dynamic modeling of future carbon emissions, making it more adaptable to potential changes in the building sector. Although several studies have concentrated on analyzing dynamic carbon trajectories and peaks associated with building decarbonization, to the best of our knowledge, limited works have investigated how to dynamically optimize provincial operational carbon emission allocations under carbon neutrality constraints by integrating static and dynamic scenario analyses.

Therefore, to fill in the gaps identified above, this work creates a top-down framework to evaluate carbon trajectories in operational commercial buildings up to 2060. Through Monte Carlo simulation, scenario analysis is conducted to assess carbon peak values and the corresponding peaking year, thereby optimizing carbon allocation schemes both nationwide and provincially. The main contributions of this work are summarized below:

**This work pioneers dynamic assessments of carbon trajectories and the corresponding carbon peaks in the operational phase of commercial buildings in China through 2060.** It

employs the proposed top-down emission model to forecast carbon trajectories and adopts scenario analysis based on the Monte Carlo simulations to dynamically model carbon peaks both nationwide and provincially. Firstly, this study utilizes the Kaya identity to build a top-down emission model under the BAU scenario and outlines a low-emission trajectory for future commercial buildings aligned with China's 2060 carbon neutrality goals under the decarbonization scenario. Additionally, Monte Carlo simulations are applied to dynamically and stochastically model future provincial carbon emissions within the constraints of carbon neutrality targets, aiming to optimize both the carbon peak values and the corresponding peaking year of commercial buildings in each province.

**This work is the first to develop a dynamic provincial-level carbon emission allocation strategy to optimize carbon allocation within commercial building operations.** By considering the initial allocation of building carbon emissions in the static BAU scenario and dynamic carbon peaks for each province via Monte Carlo simulation of future provincial building carbon emissions, this study delves into the optimization of carbon allocation strategies for each province through the coupling of static and dynamic simulations, aiming to fully leverage the decarbonization potential of commercial buildings in each province and facilitate the pace of carbon neutrality in the building sector by mid-century.

## 3. Materials and methods

Within commercial building operations, this section outlines the modeling process of the emission model and the optimized carbon allocation strategy. Section 3.1 introduces the top-down carbon emission framework developed using Kaya identity. Section 3.2 details the dynamic scenario analysis conducted via Monte Carlo simulation. Section 3.3 describes the data sources used in this study.

### *3.1. Top-down emission model*

The Kaya identity [47] was employed to develop a top-down operational carbon emission model to assess carbon emissions and analyze carbon peaks in commercial buildings both provincially and nationwide. The key parameters identified in representative studies [48], including population size ($P$), per capita floor area ($f_c$), energy intensity ($e_c$), and energy-related carbon intensity ($K_c$) of commercial buildings, were incorporated into the model. The total operational carbon emissions of commercial buildings ($C_c$) were modeled via Eq. (1):

$$C_c = P \cdot f_c \cdot e_c \cdot K_c \quad (1)$$

The parameter $K_c$ in Eq. (1) was calculated via Eq. (2):

$$K_c = K_{electricity} \cdot R_{c-electricity} + (1 - R_{c-BIPG}) \cdot (K_{coal} \cdot R_{c-coal}) + K_{gas} \cdot R_{c-gas} \quad (2)$$

Where $K_{electricity}$, $K_{coal}$, and $K_{gas}$ represent the carbon emission factors of electricity generation, coal, and natural gas, respectively. $R_{c-BIPG}$, $R_{c-coal}$, and $R_{c-gas}$ denote the proportions of self-generated energy, coal, and natural gas used in commercial buildings, respectively, and $R_{c-electricity}$ represents the end-use electrification rate of commercial buildings.

### *3.2. Dynamic scenario simulations*

Scenario analysis is an effective approach for revealing how emissions might evolve under different strategic pathways, thereby optimizing carbon allocations and guiding policy decisions to achieve long-term emission reduction goals. By adjusting key parameters—such as economic growth, energy efficiency, and technological advancements [49]—this method enables the exploration of potential carbon emission trajectories under the impact of different variables. In this study, two static scenarios, including the decarbonization and BAU scenarios, were established for China's commercial building operations. The BAU scenario serves as a baseline, reflecting the continuation of current trends in socioeconomic and technological development, without significant policy changes or innovations beyond those currently in place. The model parameters in Eq. (1) were quantitatively defined and adjusted based on historical data and official projections sourced from national statistical records and provincial energy reports. On the other hand, the decarbonization scenario assumes a more ambitious path through adjusting the potential degrees of parameter variations in the top-down emission model [see Eq. (1)], incorporating significant advancements in clean energy technologies, enhanced building efficiency, and increased adoption of renewable energy sources.

However, static scenario analysis has its limitations due to its deterministic nature, which fails to account for uncertainties in future parameter variations and potential extreme events. To address this, a dynamic model that incorporates these uncertainties through Monte Carlo simulation [50] was extended from the static emission model. The specific steps for the dynamic scenario simulation to assess the operational carbon emission peaks and optimize provincial carbon allocations were outlined as follows:

First, random parameters $\omega_i \sim N(0,\sigma_i{}^2)$, $i = P,\ f_c,\ e_c,\ K_c$ were assumed to quantify the uncertainty of parameters in the proposed emission model via Eqs. (3)-(6), thereby simulating potential parameter changes in the commercial building sector.

$$\frac{P}{(Static)} \xrightarrow[]{\substack{Random\\parameter}} \frac{P \cdot (1+\omega_P)}{(Dynamic)}, \omega_P \sim N(0,\sigma_P{}^2) \tag{3}$$

$$\frac{f_c}{(Static)} \xrightarrow[]{\substack{Random\\parameters}} \frac{f_c \cdot (1+\omega_{f_c})}{(Dynamic)}, \omega_{f_c} \sim N\left(0,\sigma_{f_c}{}^2\right) \tag{4}$$

$$\frac{e_c}{(Static)} \xrightarrow[]{\substack{Random\\parameters}} \frac{e_c \cdot (1+\omega_{e_c})}{(Dynamic)}, \omega_{e_c} \sim N\left(0,\sigma_{e_c}{}^2\right) \tag{5}$$

$$\frac{K_c}{(Static)} \xrightarrow[]{\substack{Random\\parameter}} \frac{K_c \cdot (1+\omega_{K_c})}{(Dynamic)}, \omega_{K_c} \sim N\left(0,\sigma_{K_c}{}^2\right) \tag{6}$$

By defining the variance $\sigma_i{}^2$ of each random parameter $\omega_i$ ($i = P,\ f_c,\ e_c,\ K_c$), the potential variations in each parameter in the emission model, considering uncertainty, can be quantified. A more detailed definition of parameter distributions, including mean values and variances, is provided in Appendix A.

Then, the dynamic model [see Eq. (7)], which incorporates quantitative uncertainty analysis for each parameter from 2021 to 2060, was transformed.

$$\frac{C|_T}{(Dynamic)} = \frac{C|_T}{(Static)} \cdot \left(1 + \omega_c \cdot \frac{T - 2020}{2060 - 2020}\right), \omega_c \sim N(0, {\sigma_c}^2) \quad (7)$$

To enhance the accuracy of the simulation results, a total of $100{,}000$ Monte Carlo simulations were performed for dynamic scenario simulation analysis through random sampling of the defined parameter distributions, ensuring the reliability of the final carbon peaks and the corresponding peaking time. Within China's commercial building sector, the statistical simulation results were fitted to derive the potential variation ranges of the future carbon peaks and peaking time in the BAU and decarbonization scenarios. The statistical uncertainties (expressed as $\pm \sigma_c$) were calculated based on the standard deviations (SD) of the carbon peak results across all Monte Carlo simulations, reflecting the variability introduced by the distributions of key parameters. Subsequently, provincial carbon allocations were optimized based on these simulation results, guided by the decarbonization potential maximum. These uncertainties also provided critical insights for adaptive, evidence-based policymaking. For example, positive deviations ($+\sigma_c$) may necessitate stronger decarbonization efforts, while negative deviations ($-\sigma_c$) may highlight opportunities for resource optimization or timeline adjustments.

### *3.3. Data sources*

This study covered data from 30 provinces within China's commercial building sector. However, it excluded Tibet, Hong Kong, Macau, and Taiwan, as relevant data for these regions were unavailable. National and provincial historical energy use data from 2000-2020 were collected from the *China Energy Statistical Yearbook* (available at https://www.stats.gov.cn/english/). The carbon emission factors, provided by the Intergovernmental Panel on Climate Change (available at https://www.ipcc.ch/data/), were used to calculate the historical $CO_2$ emissions by multiplying the recorded energy use. Additionally, data on population prospects were collected from the United Nations World Population Prospects (available at http://population.un.org/wpp/). Data on operational energy and carbon within commercial buildings were sourced from the Global Building Emissions Database (GLOBE, available at http://globe2060.org/).

## 4. Results

### *4.1. Prospective carbon trajectories in commercial building operations nationwide*

To tackle the primal issue outlined in Section 1, it is essential to initially depict the emission pathways within China's commercial buildings. Under the BAU and decarbonization scenarios, the static emission trajectories were simulated, represented by the dark red and yellow lines in Fig. 1, respectively. Additionally, considering the impact parameter uncertainties in Eq. (1), the prospective ranges of emission changes were also included (illustrated by the blue bands in Fig. 1). These ranges were determined through dynamic scenario analysis by assigning prior probabilities to the model parameters and correspond to -1, -2, and -3 standard deviations, representing probabilities of 32.6%, 47.8%, and 49.9%, respectively, reflecting projected decarbonization emission trends.

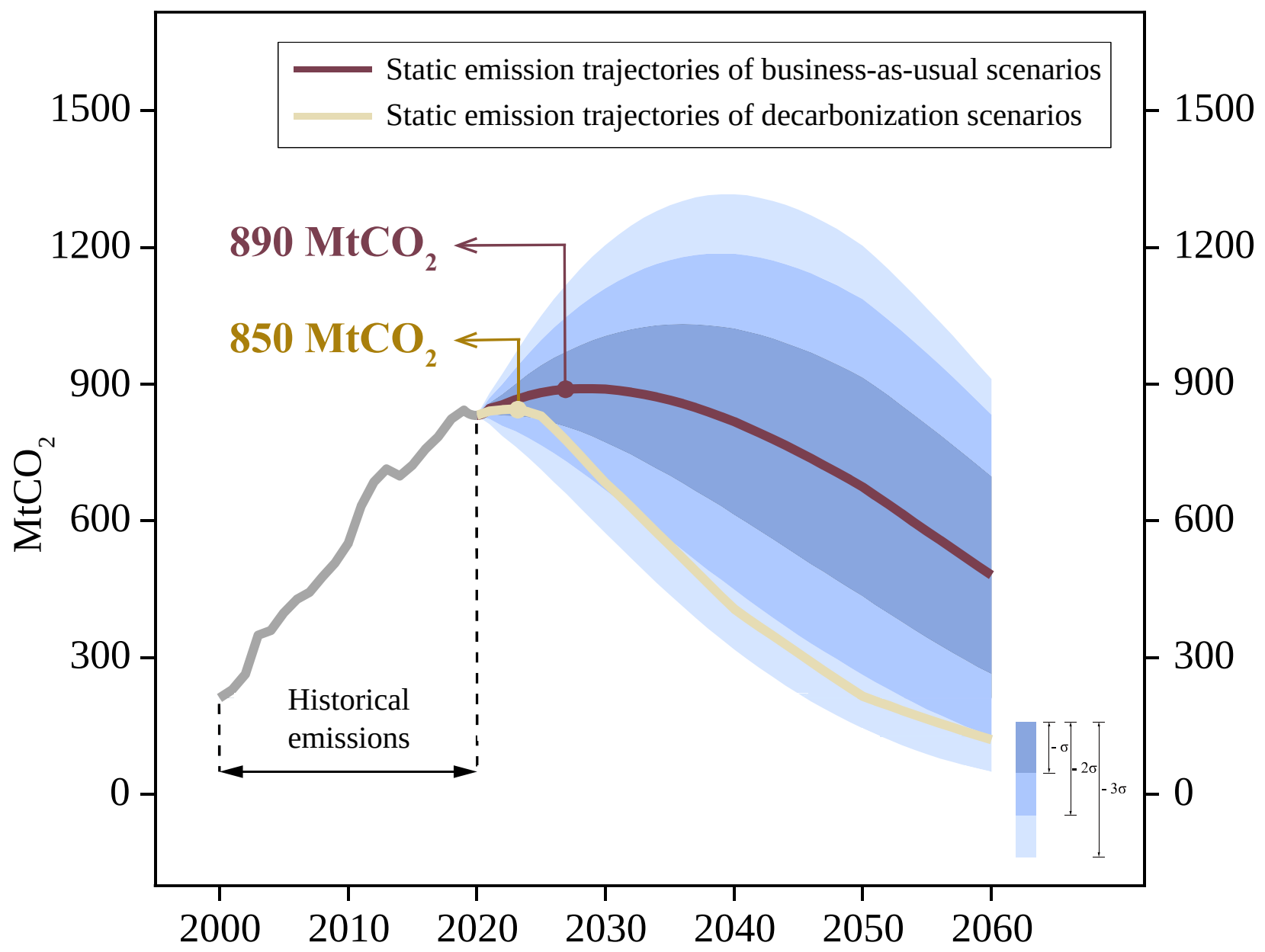

**Fig. 1.** Carbon trends and projected emission changes in the operations of commercial buildings in China through 2060.

Future emissions would exhibit an inverted U-shape under both the BAU and decarbonization scenarios, peaking at 890 megatons of carbon dioxide ($MtCO_2$) in 2028 under the BAU scenario and at 850 $MtCO_2$ before 2025 under the decarbonization scenario. This indicates that China's commercial building sector was projected to reach its peak before the 2030 target under both scenarios. Specifically, carbon emissions were forecasted to peak before 2025 under the decarbonization scenario, characterized by a greater proportion of clean energy in the mix of electricity consumed and improved building energy efficiency.

Fig. 2 shows the dynamic simulation results on the basis of 100,000 Monte Carlo simulations for energy and carbon peaks under the BAU scenario in commercial building operations. In Fig. 2 a and b, the red and blue distributions depict the prospective levels of energy use and carbon emissions, respectively, whereas the green distributions indicate the corresponding peaking time of commercial buildings.

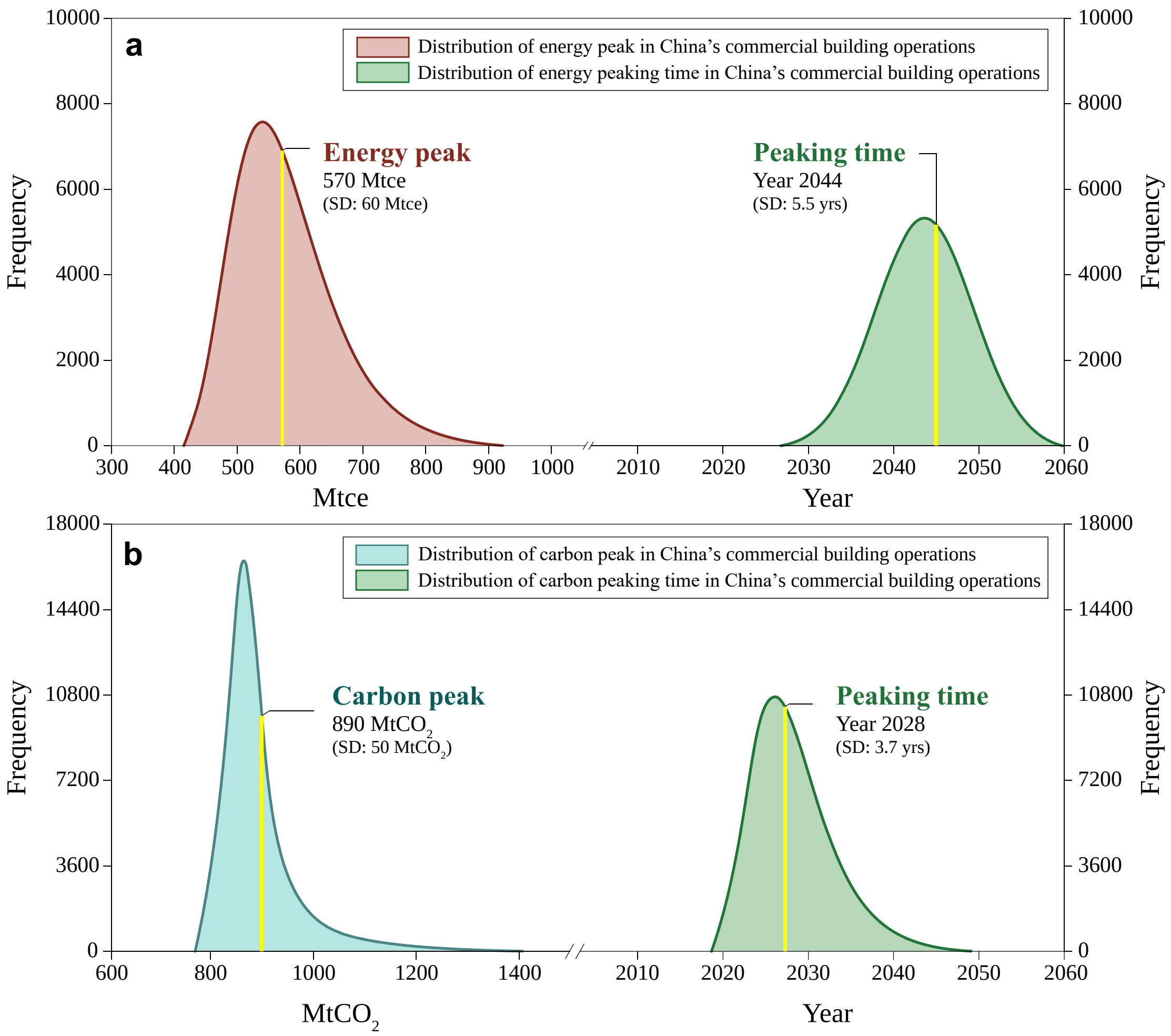

**Fig. 2.** Distributions of (a) the operational energy peak and peaking time and (b) the corresponding carbon peak and peaking time within the Chinese commercial buildings.

In Fig. 2 a, considering the uncertainty (with a 95% confidence level, ± one standard deviation), China was projected to reach its energy peak in commercial building operations by 2044 [± 5.5 years (yrs)], with a peak value of 570 (± 60) megaton of coal equivalent (Mtce). This highlights the significance of commercial building operations as a pivotal factor contributing to the early peak of energy use within the building sector. With respect to carbon emissions, as presented in Fig. 2 b, the nationwide operational emissions would peak at 890 (± 50) $MtCO_2$, with the peak expected at approximately 2028 (± 3.7 yrs). These findings suggest that commercial buildings in China could achieve carbon emission peaks before 2030, potentially facilitating the achievement of the carbon peak target within commercial buildings.

Section 4.1 outlines the nationwide emission trends and projected decarbonization pathways within commercial building operations, depicting the energy and emission peak distributions. Thus, the primary issue outlined in Section 1 has been solved.

### *4.2. Carbon peaks of commercial building operations across 30 provinces*

At the provincial scale, static scenario analyses of carbon emission peaking time and corresponding peak values within China's commercial building sector across provinces under the BAU and decarbonization scenarios up to 2060 are shown in Fig. 3, with panels a and b illustrating the emission peaks under the BAU scenario and decarbonization scenario, respectively.

In Fig. 3 a, significant differences in carbon peaks were observed among provinces under the BAU scenario, reflecting diverse carbon trajectories in the peaking time of commercial building operations. Notably, Shandong stood out with the highest carbon emission peak value at 68.9 $MtCO_2$, while Ningxia presented the lowest peak value at 5.9 $MtCO_2$. Regarding the corresponding peaking time, provinces were on track to reach their peaks as late as 2047. The regional analysis further highlights these variations: in North China, Shanxi, and Inner Mongolia were anticipated to peak first, with Hebei following in 2027. Moreover, in Northeast China, Liaoning and Jilin were expected to peak earlier, while Heilongjiang was forecasted to reach its peak in 2030. Across East China, Shanghai was projected to peak earliest, in contrast to Jiangxi's later peak in 2039. Similar disparities emerged in other regions, such as Central, South, Southwest, and Northwest China, with varying carbon peaks and the corresponding peaking time.

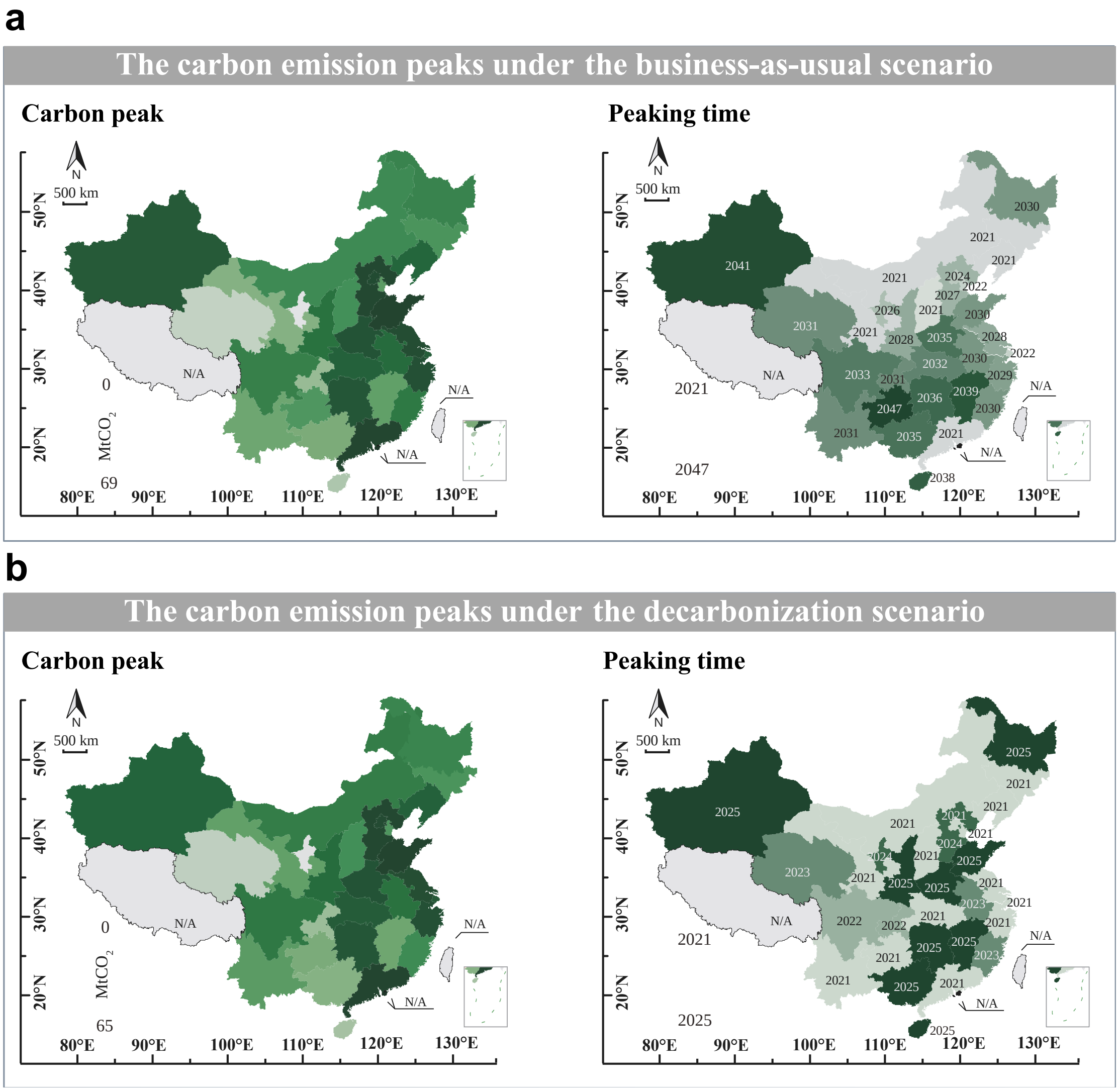

**Fig. 3.** The operational carbon peaks of provincial commercial buildings under (a) the BAU and the (b) decarbonization scenarios up to 2060.

Similar to the BAU scenario, carbon peak values varied significantly across provinces, highlighting the diverse trajectories of carbon emission timelines in commercial buildings under the decarbonization scenario. Carbon emissions across provinces exhibited disparities in peaking time, with some provinces reaching their peaks as early as 2021, whereas others were projected to peak as late as 2025. A regional analysis revealed different patterns: provinces in North China were expected to hit their peaks relatively early compared with other regions. In Northeast China, Heilongjiang's emissions were projected to peak in 2025, which contrasts with the earlier peaking time of 2021 in the other two Northeast provinces. Similarly, in East China, half of the provinces were anticipated to peak before 2030. In Central China, Hubei was projected to peak earliest, while the other two provinces were projected to peak in 2025. In South China, Guangdong was expected to peak as early as 2021, with the other two provinces expected to peak in 2025. In Southwest China, Guizhou and Yunnan were projected to peak in 2021, followed closely by Chongqing and Sichuan following in 2022. Finally, in Northwest China, Gansu was expected to peak relatively early. At the provincial level, the highest carbon peaks in commercial buildings were observed in Shandong with 65.4 $MtCO_2$, Guangdong with 61.4 $MtCO_2$, and Hebei with 58.6 $MtCO_2$. The provinces with the lowest emission peak values were Ningxia with 5.8 $MtCO_2$, Qinghai with 6.2 $MtCO_2$, and Hainan with 7.2 $MtCO_2$. Notably, Shandong's carbon emission peak value in commercial building operations was 11 times higher than that of Ningxia.

The notable disparities in carbon peaking time and corresponding peak values among provinces are primarily due to the unequal development in terms of urbanization, economic levels, and the adoption of sustainable, energy-efficient technologies. Although commercial building floor areas are continually expanding in each province, notable discrepancies exist in the growth rates of existing building floor areas and the per capita areas of commercial buildings. For example, economically advanced and highly urbanized provinces typically have a higher concentration of commercial buildings, whereas economically underdeveloped regions tend to have lower floor areas and densities of commercial buildings. Additionally, the adoption of sustainable, energy-efficient technologies varies across provinces, with some provinces implementing more effective energy-saving measures earlier, consequently leading to a reduction in carbon emission levels.

Overall, within commercial buildings, the detailed results of the carbon peaks reveal significant variations across provinces, providing an analysis of the underlying reasons. Thus, the majority of the second problem displayed in Section 1 has been addressed.

## 5. Discussion

The goal of realizing carbon neutrality by mid-century in commercial buildings is challenging, as evidenced by the static operational emission trajectories and the notable disparities among provinces in reaching carbon emission peaks, as shown in Section 4. This section further explores the dynamic emission reduction potential across provinces under the decarbonization scenario and optimizes provincial carbon allocations within commercial buildings on the basis of the decarbonization potential maximum, thereby promoting the national carbon neutrality goal.

### *5.1. Chances of commercial buildings shifting toward a decarbonization scenario*

Integrating static emission modeling with dynamic simulations provides a comprehensive analysis of carbon peak scenarios. First, the static emission model was used to assess the carbon peaking time and corresponding emission peak range for commercial building operations nationwide under the decarbonization scenario. Subsequently, dynamic emission simulations of future operational building carbon emissions nationwide under the BAU scenario were conducted via Monte Carlo simulations to assess the probability of achieving the corresponding emission peak range under the decarbonization scenario. The probability analysis results, which illustrate carbon peaks of the decarbonization scenario in future operations nationwide under the BAU scenario, are illustrated in Fig. 4.

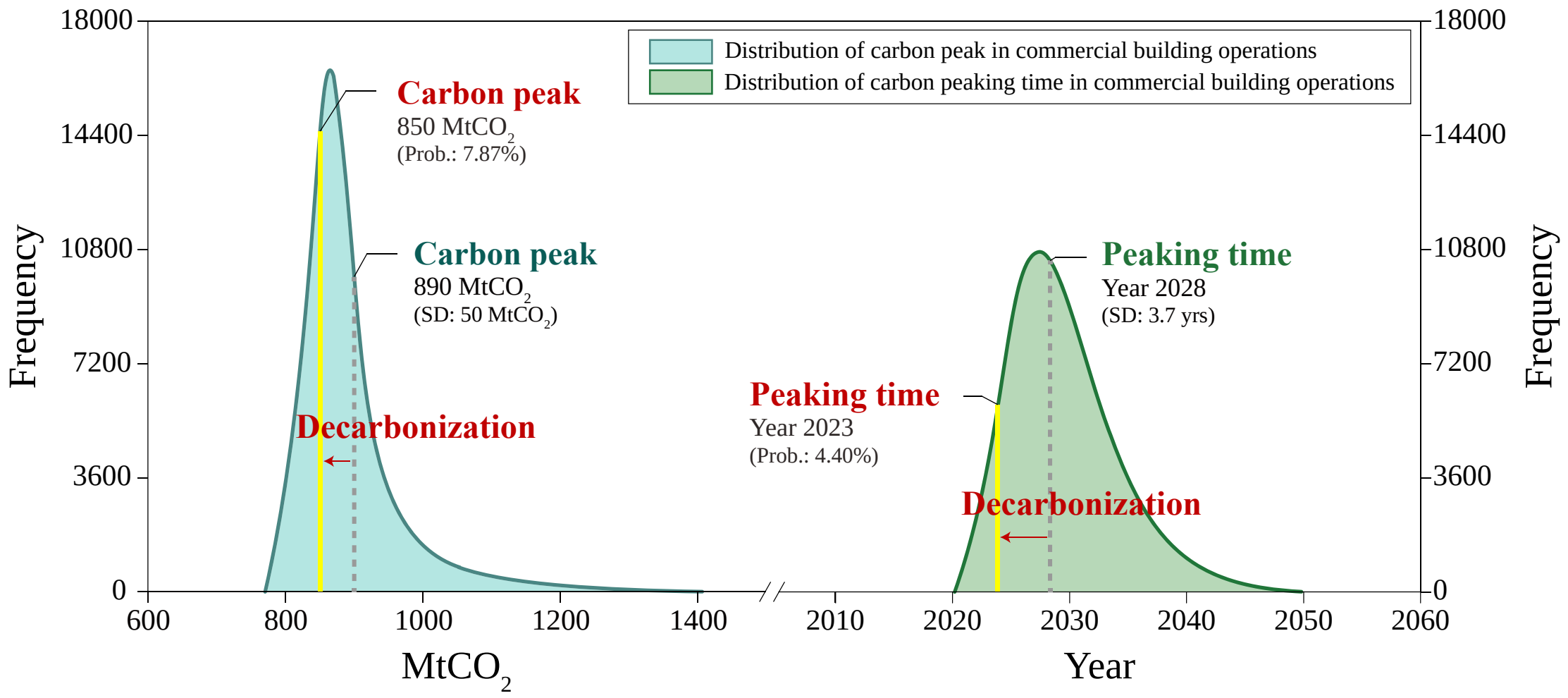

**Fig. 4.** Distributions of the carbon peak and peaking time probabilities under the decarbonization scenario in the operation of commercial buildings nationwide.

Considering the uncertainties, the nationwide emissions under the BAU scenario would peak at 890 (± 50) $MtCO_2$, with the peaking time expected at approximately 2028 (± 3.7 yrs). Under the decarbonization scenario, there was a 4.40% probability that commercial buildings would reach the carbon peak by 2023, nearly five years earlier. The probability of reaching the carbon emission peak interval was 7.87% under the decarbonization scenario, with the corresponding estimated carbon emission peak value at 850 $MtCO_2$, indicating a reduction of 40 $MtCO_2$ in the carbon emission peak compared with the BAU scenario.

### *5.2. Provincial allocations of commercial building emissions towards carbon neutrality*

With a probability of 7.87% of achieving emission peaks under the decarbonization scenario, dynamic scenario simulations of the carbon peak probability and peak range for each province are illustrated in Figs. 5 and 6, which include the carbon peak values and the corresponding peaking time across provinces under the BAU scenarios (depicted in red text) and the decarbonization scenario (highlighted in dark text) in Figs. 5 and 6, respectively. Additionally, guided by the principle of the decarbonization potential maximum, optimal allocation schemes for commercial buildings in each province were further analyzed.

Regarding the future carbon emission peaking time and decarbonization targets across provinces, as presented in Fig. 5, the peaking time ranged from as early as 2021 to as late as 2046. Provinces with earlier peaking time included Jining in 2021 (± 1.2 yrs), Guangdong and Inner Mongolia in 2022 (± 1.4 yrs), and Liaoning, Shanxi, and Shanghai in 2023 (± 2.1, ± 3.1 and ± 2.5 yrs, respectively). Conversely, the provinces with the latest peaking time were Jiangxi in 2039 (± 4.2 yrs), Hainan and Xinjiang in 2040 (± 5.6 and ± 3.8 yrs), and Guizhou in 2046 (± 4.8 yrs). Compared with the BAU scenario, each province needs to reach its carbon emission peak earlier to achieve the decarbonization target. At the provincial level, several provinces, including Guizhou (in 10 yrs), as well as Sichuan, Yunnan, and Henan (in 9 yrs), particularly should endeavor to advance the achievement of carbon peaks to promote the decarbonization target.

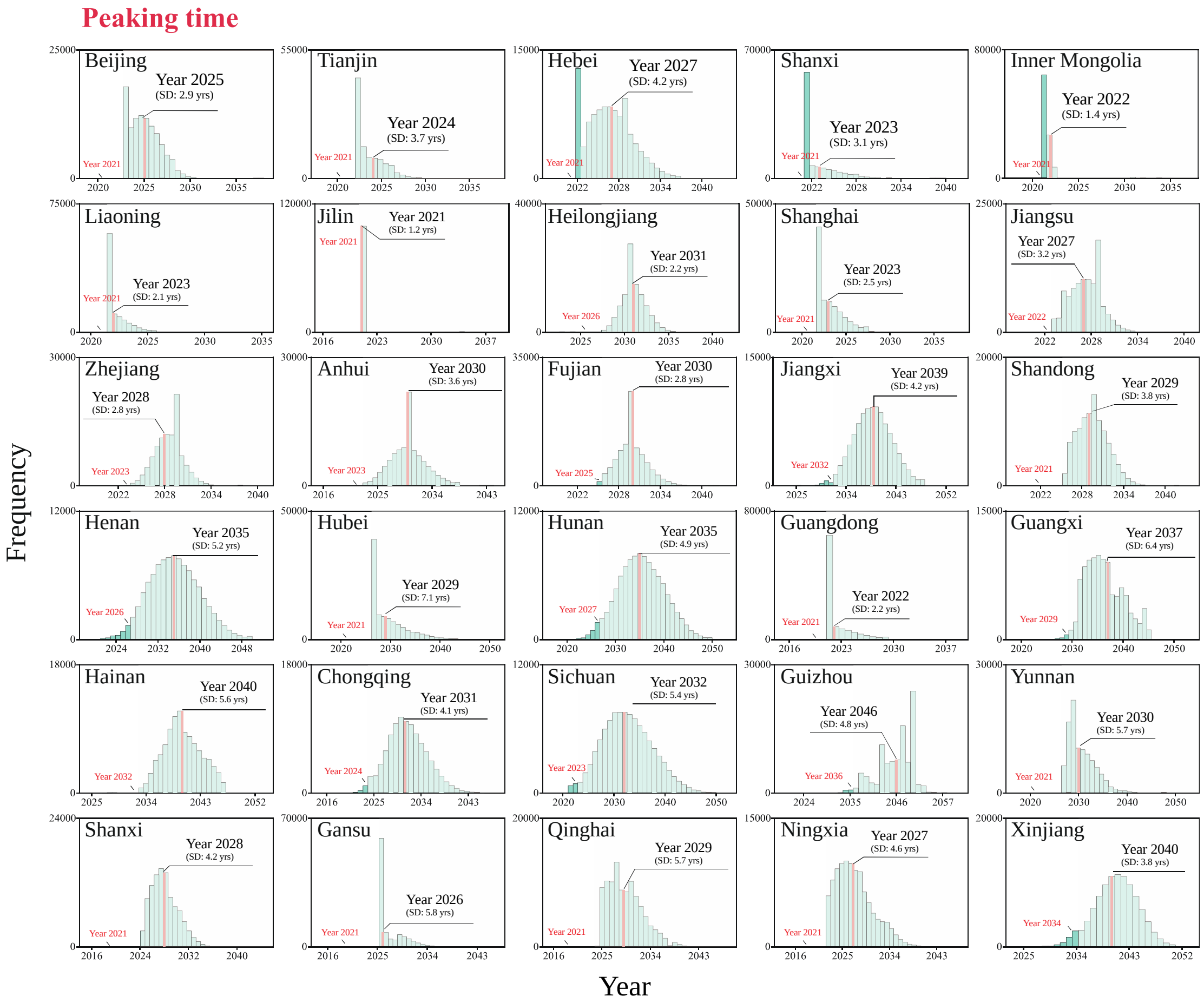

**Fig. 5.** Distributions of operational carbon peaking time in commercial buildings across 30 provinces in China.

According to Fig. 6, significant variations in future carbon emission peaks were evident across provinces, considering uncertainties under the BAU scenario. Specifically, the top three provinces in respect of future carbon peaks were Shandong (69.6 ± 4.0 $MtCO_2$), Guangdong (62.5 ± 1.0 $MtCO_2$), and Hebei (61.8 ± 2.8 $MtCO_2$). Conversely, the last three provinces were Ningxia (6.0 ± 0.3 $MtCO_2$), Qinghai (6.6 ± 0.4 $MtCO_2$), and Hainan (9.7 ± 1.2 $MtCO_2$). Notably, the emission peak value in Shandong was approximately 11 times greater than that in Ningxia. Variations in population density, energy consumption structure, economic activity, substantial commercial building stock, etc., across provinces contributed to disparities in emissions. For example, Shandong, one of China's most industrialized provinces, had a high concentration of energy-intensive industries and substantial commercial building stock, leading to increased energy consumption and emissions. On the other hand, Guangdong, as an economic powerhouse with rapid urbanization, boasted a large and expanding base of commercial activities, which significantly increased energy demand in the building sector [51]. Additionally, both provinces exhibited above-average population densities and GDP levels, further contributing to their elevated emissions.

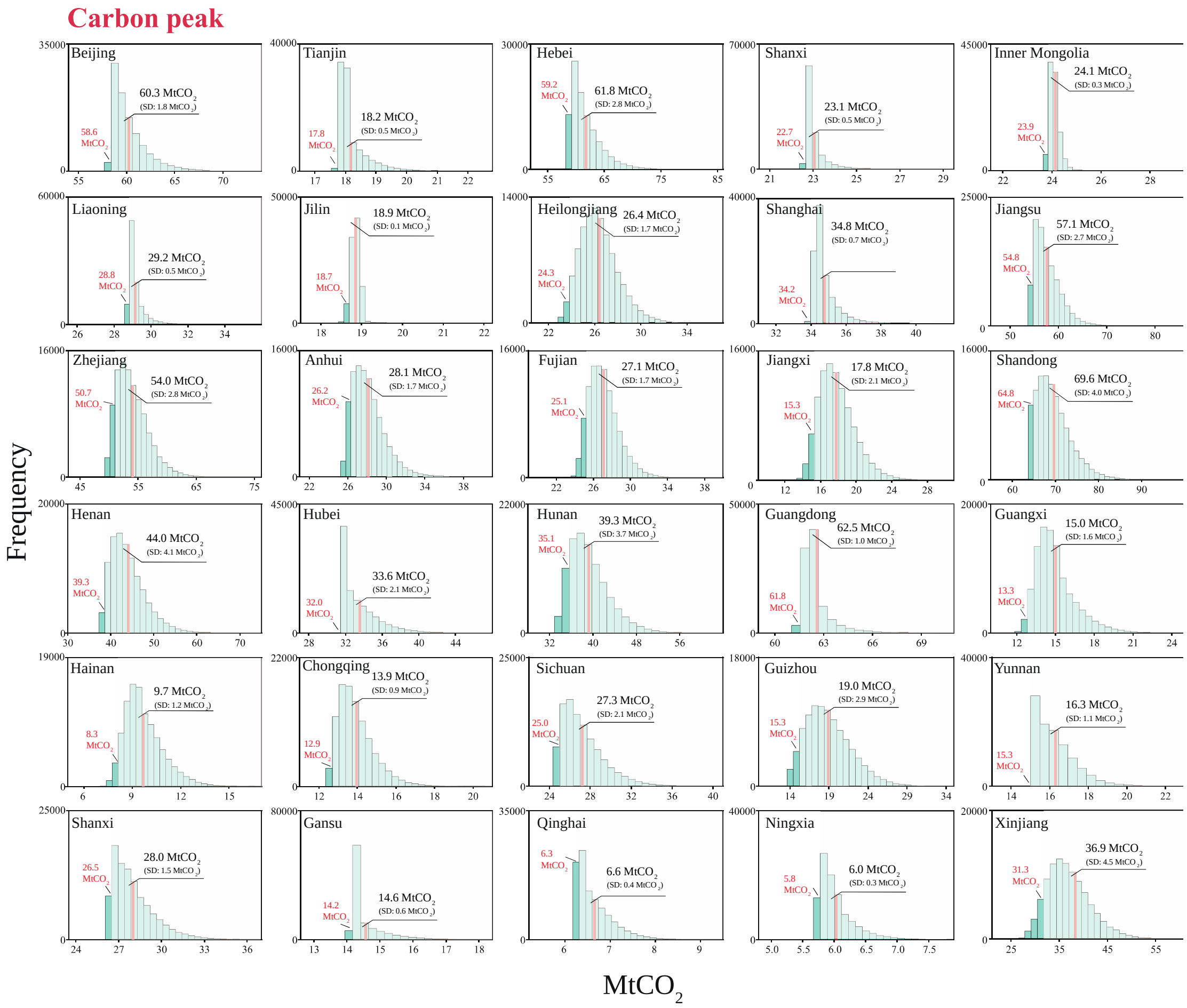

**Fig. 6.** Distributions of operational carbon peaks in commercial buildings across 30 provinces in China.

To mitigate regional emission disparities, each province needs to reduce its carbon emissions according to the estimated results under the BAU and decarbonization scenarios. Guided by the decarbonization potential maximum, the optimal strategy for allocating emissions across provinces was determined. At the provincial level, the top three provinces requiring emission reductions were Xinjiang (5.6 $MtCO_2$), Shandong (4.8 $MtCO_2$), and Henan (4.7 $MtCO_2$), while the last three provinces were Jilin (0.2 MtCO2), Inner Mongolia (0.3 $MtCO_2$), and Ningxia (0.3 $MtCO_2$), with the highest reduction amount being 28 times the lowest. Regionally, East China should undertake the greatest responsibility for emission reduction, with a total reduction requirement of 18.1 $MtCO_2$ across seven provinces. In Central China, the average reduction responsibility for the three provinces was the highest at 2.6 $MtCO_2$. In contrast, Northeast China would have the smallest total reduction responsibility of 2.7 $MtCO_2$ across the three provinces. To achieve these reductions, policy and technological interventions should be tailored to the specific characteristics of each province. For example, Xinjiang and Shandong could prioritize integrating renewable energy sources and optimizing the energy storage system [52], such as solar and wind power, into their energy mix to reduce reliance on coal and other fossil fuels. Shandong, given its significant heating demand, could also focus on transitioning from coal-based heating to cleaner alternatives, such as district heating powered by renewable energy or high-efficiency heat pumps. Provinces with high building operation emissions, such as Henan, should prioritize improving energy efficiency through retrofitting existing buildings and upgrading heating, ventilation, and air conditioning systems, while addressing the water footprint of the construction sector to tackle environmental challenges [53]. By combining region-specific strategies with broader national policies, these measures can help achieve equitable and effective emission reductions across provinces.

Overall, the second and the third problems presented in Section 1 are entirely solved.

### *5.3. Policy implications*

The commercial building sector plays a crucial role in achieving net-zero emission goals. However, the traditional “one-size-fits-all” approach and locally self-declared emission reduction strategies are insufficient in addressing the complexities of regional disparities. Achieving meaningful progress requires a unified national framework that establishes overarching targets and guidelines for the sector, complemented by region-specific emission reduction policies. Provinces should adopt differentiated policies tailored to their unique carbon emission profiles, economic conditions, and technological capacities. This dual-layered approach ensures alignment with national carbon neutrality goals while enabling each province to make substantial contributions to the decarbonization of commercial buildings.

To effectively address carbon emissions in the commercial building sector, policymakers should implement a unified framework mandating higher energy efficiency standards for new construction, including the full adoption of green building standards and compulsory ultra-low-energy design, such as urban green roofs [54] and vehicle-to-grid with integrated energy systems [55]. Simultaneously, retrofitting existing commercial buildings is essential to optimizing energy performance, particularly for high-energy-consuming structures. Key measures include upgrading exterior walls, windows, and equipment, as well as integrating smart technologies for real-time energy monitoring and dynamic management. Furthermore, increasing electrification rates within commercial buildings by incentivizing the adoption of clean and renewable energy sources is vital for advancing comprehensive electrification in the building sector [56].

On the other hand, regional governments should establish green material supply chain platforms to ensure access to energy-efficient materials and advanced construction technologies. Recognizing regional differences, tailored policies are essential for achieving equitable and effective low-carbon goals. High-emission regions such as Shandong, Guangdong, and Hebei should strengthen regulations, incentivize stricter emission standards, and promote the utilization of clean energy and green materials [57]. Cold-climate areas like Heilongjiang should mandate green and ultra-low-energy standards for new buildings while expanding retrofitting programs to enhance thermal performance [58]. Resource-rich provinces such as Inner Mongolia and Xinjiang can leverage renewable energy potential, particularly solar photovoltaic [59], to steadily increase substitution rates in the commercial building sector. Other regions, such as Anhui, Hubei, Yunnan, Chongqing, etc., should prioritize strengthening low-carbon material supply chains, enforcing energy management standards, and promoting near-zero-energy demonstration projects. By implementing regionally differentiated policies and fostering interregional cooperation, policymakers can effectively advance decarbonization in China's commercial building sector, supporting net-zero and sustainable development goals [60].

The findings underscore the critical role of China's commercial building sector in the nation's broader energy transition and climate goals, aligning with its commitment to peak carbon emissions before 2030 and achieve carbon neutrality by 2060. As the world's largest emitter, China's efforts to optimize regional decarbonization strategies provide valuable insights and serve as baselines for other nations striving for carbon neutrality in global climate challenges.

## 6. Conclusions

This work proposed a top-down emission model combined with dynamic scenario analysis via Monte Carlo simulations to assess operational carbon emission trends and evaluate the corresponding operational carbon peaks in commercial buildings of China from 2000-2060. The analysis was conducted both nationwide and provincially, highlighting spatial and temporal disparities in emission trajectories. Furthermore, optimized strategies were proposed for allocating provincial-level carbon emissions on the basis of the maximum decarbonization potential, ensuring equitable and effective emission reductions across regions. Finally, the policy implications for achieving carbon neutrality in China's commercial buildings were discussed. The core findings are summarized as:

### *6.1. Core findings*

- **The simulated emission trajectories from 2000 to 2060 would exhibit inverted U-shaped curves in China's commercial building operations**. The emissions of national commercial building operations would peak in 2028 at 890 Mt$CO_2$ under the BAU scenario, and the probability of achieving carbon peaks under the decarbonization scenario stood at 7.87%, with a corresponding peak level of 850 Mt$CO_2$. Considering the uncertainty, nationwide carbon emissions would peak at 890 (± 50) Mt$CO_2$ by 2028 (± 3.7 yrs) under the BAU scenario for realizing operational decarbonization in commercial buildings. In addition, China was projected to reach its energy peak in commercial building operations by 2044 (± 5.5 yrs), with a peak value of 570 (± 60) Mtce.
- **Significant disparities in carbon emission peaks were observed across provinces and regions under both the BAU and decarbonization scenarios**. Considering the uncertainties in dynamic Monte Carlo simulations, emissions across provinces were anticipated to peak before 2046 (± 4.8 yrs). Shandong, Guangdong, and Hebei emerged as the top three provinces with the highest carbon peaks, with values of 69.6 (± 4.0) Mt$CO_2$ in 2029 (± 3.8 yrs), 62.5 (± 1.0) Mt$CO_2$ in 2022 (± 2.2 yrs), and 61.8 (± 2.8) Mt$CO_2$ in 2027 (± 4.2 yrs), respectively. Conversely, Ningxia, Qinghai, and Hainan presented the lowest carbon peaks, with 6.0 (± 0.3) Mt$CO_2$ in 2027 (± 4.6 yrs), 6.6 (± 0.4) Mt$CO_2$ in 2029 (± 5.7 yrs), and 9.7 (± 1.2) Mt$CO_2$ in

2040 (± 5.6 yrs), respectively.

- **The optimization strategy for carbon allocation revealed significant regional variation in emission reduction across provinces, with the highest reduction needs amounting to 6.7 times the lowest**. The top three provinces requiring emission reductions would be Xinjiang (5.6 $MtCO_2$), Shandong (4.8 $MtCO_2$), and Henan (4.7 $MtCO_2$), while the bottom three provinces would be Jilin (0.2 $MtCO_2$), Inner Mongolia (0.3 $MtCO_2$), and Ningxia (0.3 $MtCO_2$). In terms of regions, East China stood out with the highest emission reduction requirement of 18.1 $MtCO_2$ across seven provinces, whereas the Central China region demonstrated the highest average reduction responsibility of 2.6 $MtCO_2$ across three provinces. Conversely, the Northeast region showed the lowest total reduction responsibility of 2.7 $MtCO_2$ across three provinces.

### *6.2. Future work*

To optimize carbon emission allocation strategies in the commercial building sector and progress toward China's net-zero goals, several research gaps warrant further investigation. While the model captures key drivers of operational carbon emissions, its simplifications may overlook other influential factors, and its parameter assumptions—defined based on historical data and current policies—carry uncertainties regarding future changes. Future work should incorporate additional variables, such as water and space heating, as well as near-zero-energy building technologies, to assess the long-term decarbonization potential of building operations and clarify the roadmap for deep decarbonization. Moreover, integrating renewable energy sources, such as solar photovoltaics and wind energy, into building operations should be prioritized to enhance energy efficiency and decarbonization efforts. Advanced technologies, including smart building systems and energy storage solutions, should also be explored for their potential to enable carbon monitoring, predictive maintenance, and demand-side energy management in the building sector.

## Acknowledgment

First author appreciates the National Planning Office for Philosophy and Social Science Foundation of China (24BJY129).

## Appendix

The appendix is available in the supplementary material (e-component) accompanying this submission.